\documentclass[aps,prc,letterpaper,11pt,twoside,preprintnumbers,tightenlines,superscriptaddress,showpacs,nofootinbib]{revtex4-1}
\usepackage{graphicx,times}
\usepackage{amsmath,amsfonts,amssymb}

\begin{document}

\title{An alternative explanation for the $\Theta(1540)$ pentaquark peak}

\author{A. Mart\'inez Torres}
\affiliation{
Yukawa Institute for Theoretical Physics, Kyoto University,
Kyoto 606-8502, Japan}

\author{E. Oset}
\affiliation{Departamento de F\'isica Te\'orica and IFIC, Centro Mixto Universidad de Valencia-CSIC, Institutos de Investigaci\'on de Paterna, Aptdo. 22085, 46071 Valencia, Spain.}

\preprint{YITP-10-83}

\begin{abstract}
 We have studied the $\gamma d \to K^+ K^- n p$ reaction in which the LEPS collaboration found
 a signal in the $K^+ n$ invariant mass for the claimed $\Theta(1540)$ pentaquark peak.
 Our study reveal that the procedure used at LEPS to reconstruct the $K^+n$ invariant mass generates
 an artificial strength in the $\Theta(1540)$ region and that the LEPS collaboration underestimated the background 
 in that region, such that the signal observed for the $\Theta(1540)$ is compatible with a fluctuation of $2\sigma$ over the calculated
 background.
 \end{abstract}

\maketitle


\section{Introduction}
In 1997 Diakonov and collaborators proposed the existence of a low-mass antidecuplet of pentaquarks
with spin $1/2$ and even parity \cite{diakonov}. In particular, they predicted the existence of a narrow state, around 15 MeV of width or less, with $S=+1$
close to a mass of 1530 MeV. In 2003, the baryon spectroscopy physics experimented a revolution since the LEPS collaboration reported the observation of a narrow peak in the $K^+n$ invariant mass
in the reaction $\gamma n \to K^+ K^- n$ on a $^{12} C$ target \cite{penta1} which was associated to the presence of a pentaquark state with
$S=1$, named $\Theta(1540)$, and which was considered as a possible candidate for the state predicted in \cite{diakonov}. After the LEPS experiment, many different experimental groups
started searching for evidences of that $\Theta(1540)$ in different reactions and many experiments were done: some of them, in general with poor statistics, were able to find a signal while others, with large statistics, could not, and the issue stimulated a large effort to explain the nature of the signal observed by the LEPS collaboration (see the list of references in \cite{pdg}). After a period of rest, in 2009 the LEPS collaboration made a new experiment, this time they analyzed the $\gamma d \to K^+ K^- p n$ reaction and with more statistics, confirming the existence of the $\Theta(1540)$ \cite{penta2}.  However, the CLAS collaboration studied the same reaction than in \cite{penta2} but with ten times more statistics and with complete kinematics (but excluding small angles) and did not find any signal which could be associated with the $\Theta(1540)$ \cite{clas}. It is important to stress here the detail of complete kinematics since it plays an important role in the analysis of the reaction at LEPS due to the fact that neither the proton nor the nucleon are detected and one needs some prescription to guess the momenta of these particles and then reconstruct the corresponding invariant masses. 

In this talk I will present a model which we have done to study the latest LEPS reaction. We have adapted our model to simulate the LEPS set up and see the effect that the prescription and cuts done at LEPS in the reconstruction of the $K^+ n$ and $K^- p$  can generate in their respective spectra.

\section{Formalism}
In order to simulate the $\gamma d\to K^{+}K^{-}np$ reaction studied at LEPS \cite{penta2} we consider the basic features observed from the experiment. The most important contribution comes from $\phi$ production, the elementary reaction $\gamma p\to \phi p$ and $\gamma n\to \phi n$. In our simulation,  we implement this $\phi$ production through a minimal model which incorporates the basic structure of $K^{+}K^{-}$ production correlated by the $\phi$ propagator. Another  element for consideration in the dynamics of the $\gamma d\to K^{+}K^{-}np$ reaction is the $\Lambda(1520)$ production on the proton. Since we are only interested in the shape of the $K^ - p$ distribution and the strength of the $\Lambda(1520)$, here we consider a minimal structure to describe the $\gamma p \to K^+ \Lambda(1520)\to K^+ K^- p$ which is compatible with the quantum numbers and $D$ wave character of the $\Lambda(1520)$. 
The other  important  element in our approach is the rescattering that unavoidably occurs in the reaction in deuteron. The details of the model can be found in \cite{MO1}.

The LEPS detector is a forward magnetic spectrometer and its geometry is implemented in our simulation by imposing that the angle of the kaons in the final state with respect the incident photon is not bigger than 20 degrees. 

The nucleons are not detected at LEPS, therefore, some prescription is required in order to estimate the momentum of the $p$ and $n$ in the reaction  $\gamma d \to K^{+}K^{-}np$ and determine the invariant mass of $K^{-}p$ or $K^+ n$. This is done using the minimum momentum spectator approximation (MMSA). For this purpose one defines the magnitude 
\begin{equation}
p_{pn}=p_{miss}=p_{\gamma} +p_d - p_{K^+} -p_{K^-}
\end{equation}
which corresponds to the four momentum of the outgoing $pn$ pair. From there one evaluates the nucleon momentum in the frame of reference where the $pn$ system is at rest, $\vec{p}_{CM}$.
Boosting back this momentum to the laboratory frame, we will have a minimum modulus for the momentum of the spectator nucleon when the momentum $\vec{p}_{CM}$ for this nucleon goes in the direction opposite to $\vec{p}_{miss}$. Thus, the minimum momentum, $p_{min}$, is given by 
 \begin{equation}
 p_{min} = -|\vec{p}_{CM}| \cdot \frac{E_{miss}}{M_{pn}} + E_{CM} \cdot\frac{|\vec{p}_{miss}|}{M_{pn}}
\end{equation}
where $E_{CM}=\sqrt{|\vec{p}_{CM}|^{2}+M^{2}_{N}}$ is the energy of the nucleon in the CM frame. In this case, the momentum of the other nucleon will be in the direction of the missing momentum  with a magnitude 
\begin{equation}
p_{res} = |\vec{p}_{miss}| -p_{min}	
\end{equation}
In \cite{penta2} the $M_{K^+ n}$ invariant mass for the reaction $\gamma d\to K^{+}K^{-}np$ is evaluated assuming the proton to have a momentum $p_{min}$ (actually what one is evaluating is $M_{K^+ N}$, with $N$ the non spectator nucleon). Consequently, in this prescription, the momentum of the neutron in the final state will be
\begin{equation}
\vec{p}_{n}=p_{res}\cdot\frac{\vec{p}_{miss}}{|\vec{p}_{miss}|}
\end{equation}
which is used to calculate the $M_{K^{+}n}$ invariant mass for the reaction $\gamma d\to K^{+}K^{-}np$ in \cite{penta2}.  A cut is imposed at LEPS demanding that  $|p_{min}|<$ 100 MeV. This condition is also implemented in our simulation of the process.

The contribution from the $\phi$ production at LEPS is removed considering events which satisfy that the invariant mass of the $K^{+}K^{-}$ pair is bigger than 1030 MeV and bigger than the value obtained from the following expression
\begin{equation}
1020\, \textrm{MeV} +0.09\times (E^{eff}_{\gamma}(\textrm{MeV})-2000\, \textrm{MeV})
\end{equation}
where $E^{eff}_{\gamma}$ is defined as the effective photon energy
\begin{equation}
E^{eff}_{\gamma}=\frac{s_{K^{+}K^{-}n}-M^{2}_{n}}{2M_{n}}\label{Eeff}
\end{equation}
with $s_{K^{+}K^{-}n}$ the square of the total center of mass energy for the $K^{+}K^{-}n$ system calculated using the MMSA approximation to determine the momentum of the neutron assuming the proton as spectator. In \cite{penta2} only events for which $2000$ MeV $<E^{eff}_{\gamma} $ $<2500$ MeV are considered, a condition which is also incorporated in our simulation. The $E^{eff}_{\gamma}$ of Eq. (\ref{Eeff}) with the MMSA prescription would correspond to the photon energy in the frame where the original non spectator (participant) nucleon is at rest.

\section{Results}

\begin{figure}[h!]
\centering
\includegraphics[width=0.33\textwidth]{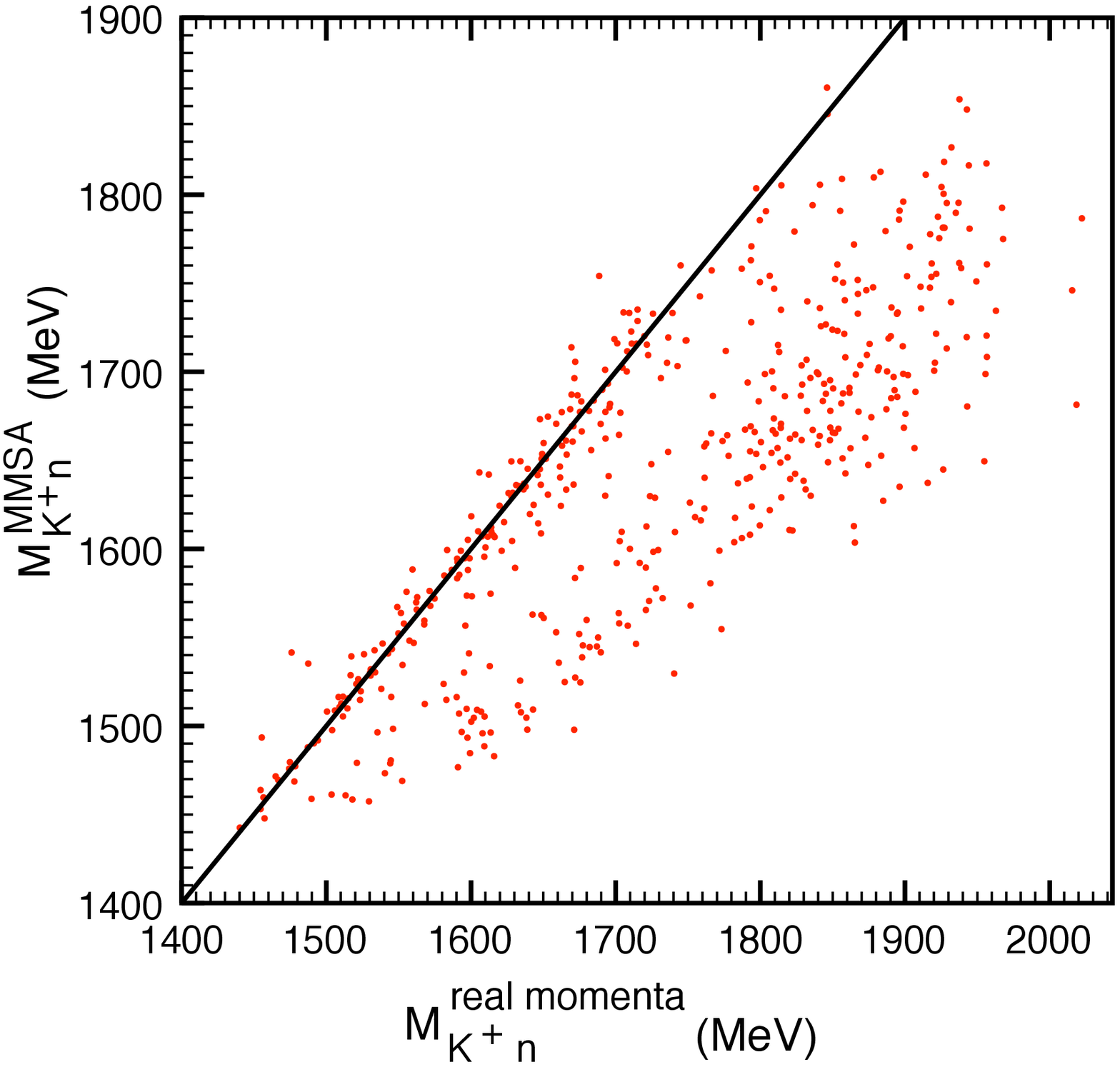}
\includegraphics[width=0.33\textwidth]{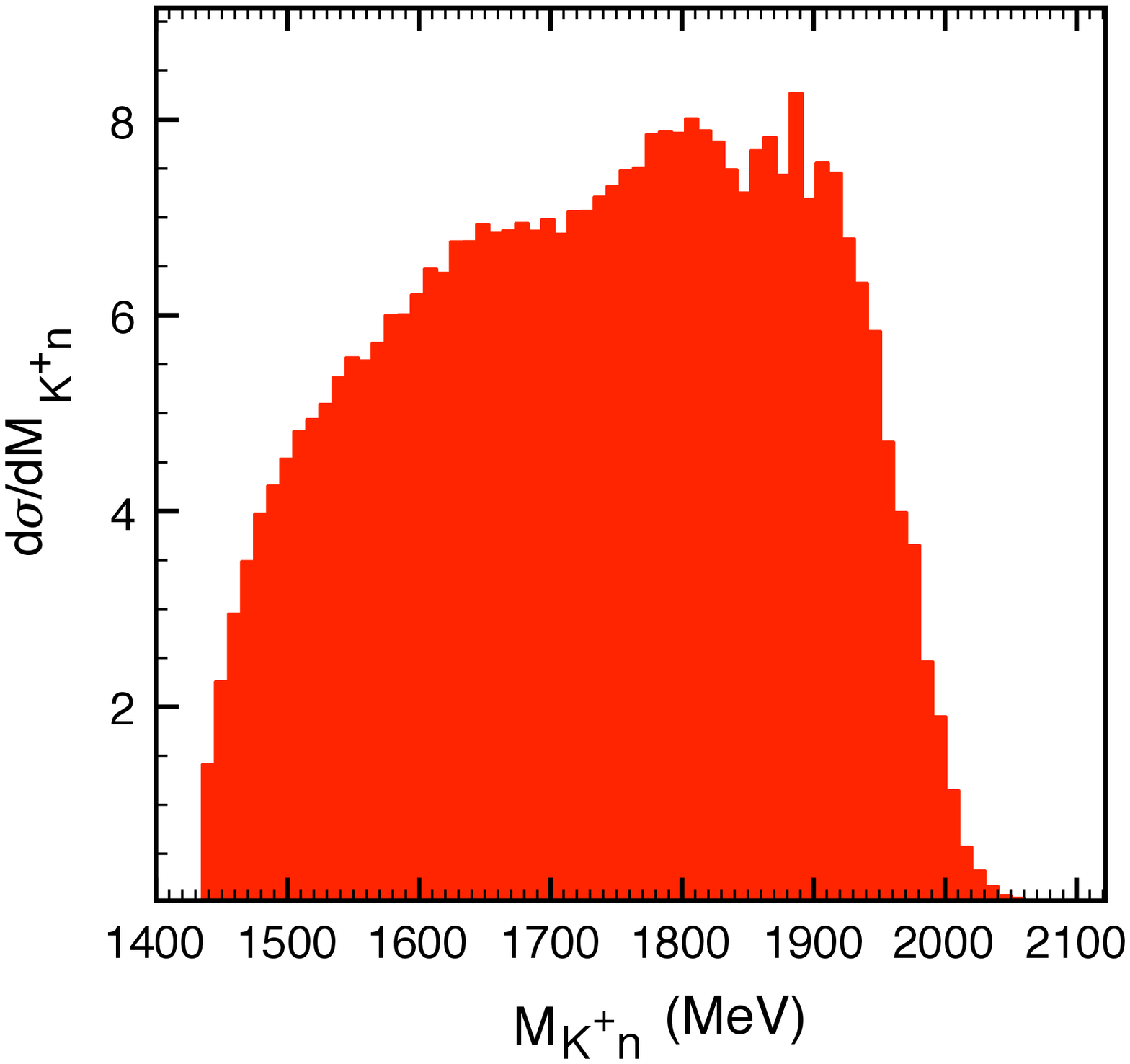}

\caption{(Left) $M_{K^{+}n}$ calculated using the MMSA prescription versus $M_{K^{+}n}$ obtained with the real momentum for the nucleons and the full model, i.e., $\phi$ production on the nucleons and $\Lambda(1520)$ production on the proton. (Right) $M_{K^{+}n}$ invariant mass distribution calculated using the real momenta and with a $\phi$ cut of $M_{K^{+}K^{-}}>1050$ MeV.}\label{Minvnaka_real}
\end{figure}

In Fig. \ref{Minvnaka_real} (Left) we show  the  $K^{+}n$ invariant mass distribution for the LEPS set up using the real momenta obtained from our Monte Carlo integral of the cross section versus the one obtained using the momenta determined with the MMSA prescription. We can distinguish two blocks of points: one of them distributed around the diagonal and another one with points scattered around the plane. This is like that  because actually the MMSA prescription reconstructs the $K^+ N$ invariant mass, where $N$ is the participant nucleon, which about half of the times is the neutron and the other half the proton. The points around the diagonal correspond to the case where the participant is the neutron.

\begin{figure}[h!]
\centering
\includegraphics[width=0.33\textwidth]{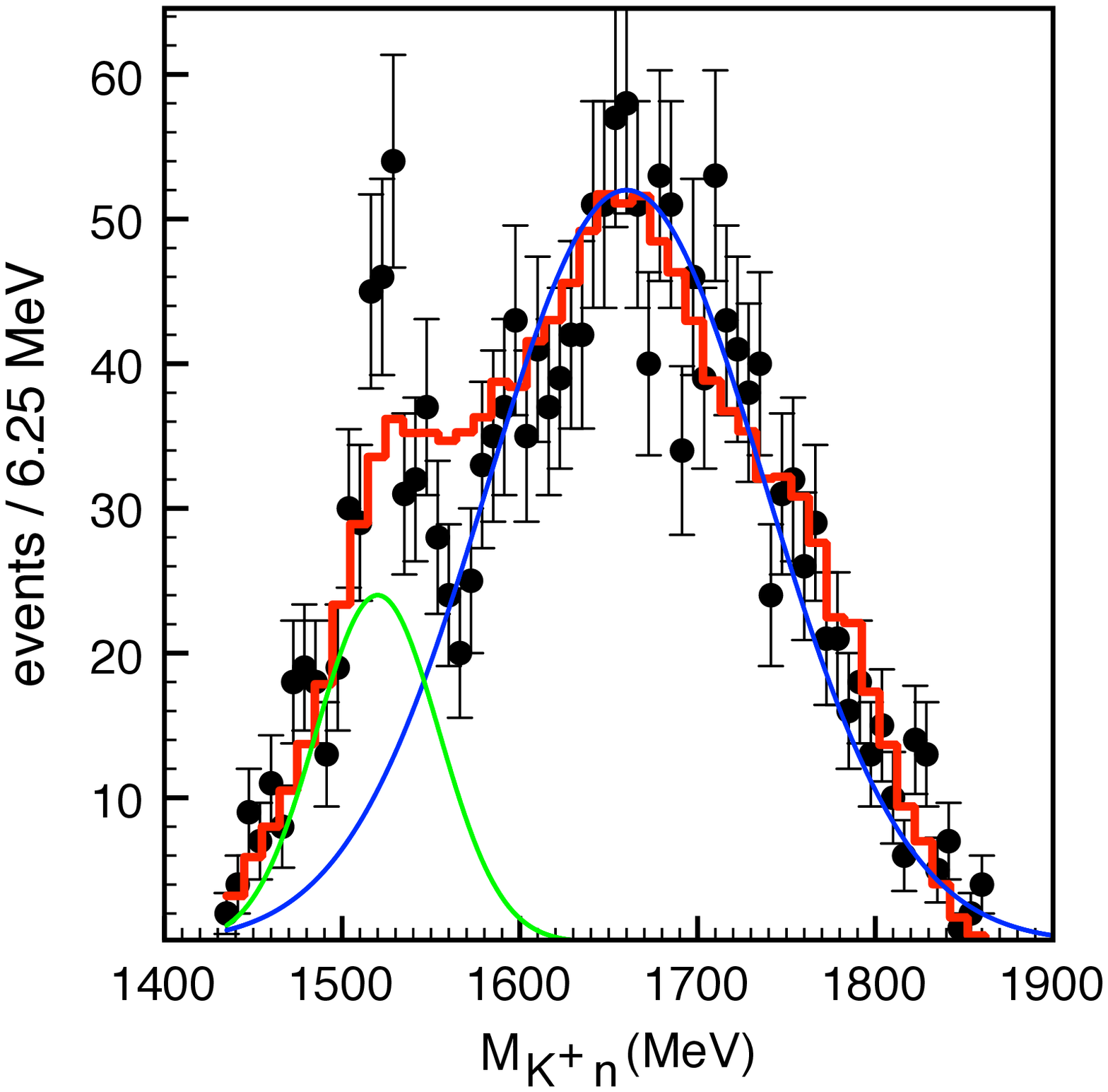}
\includegraphics[width=0.33\textwidth]{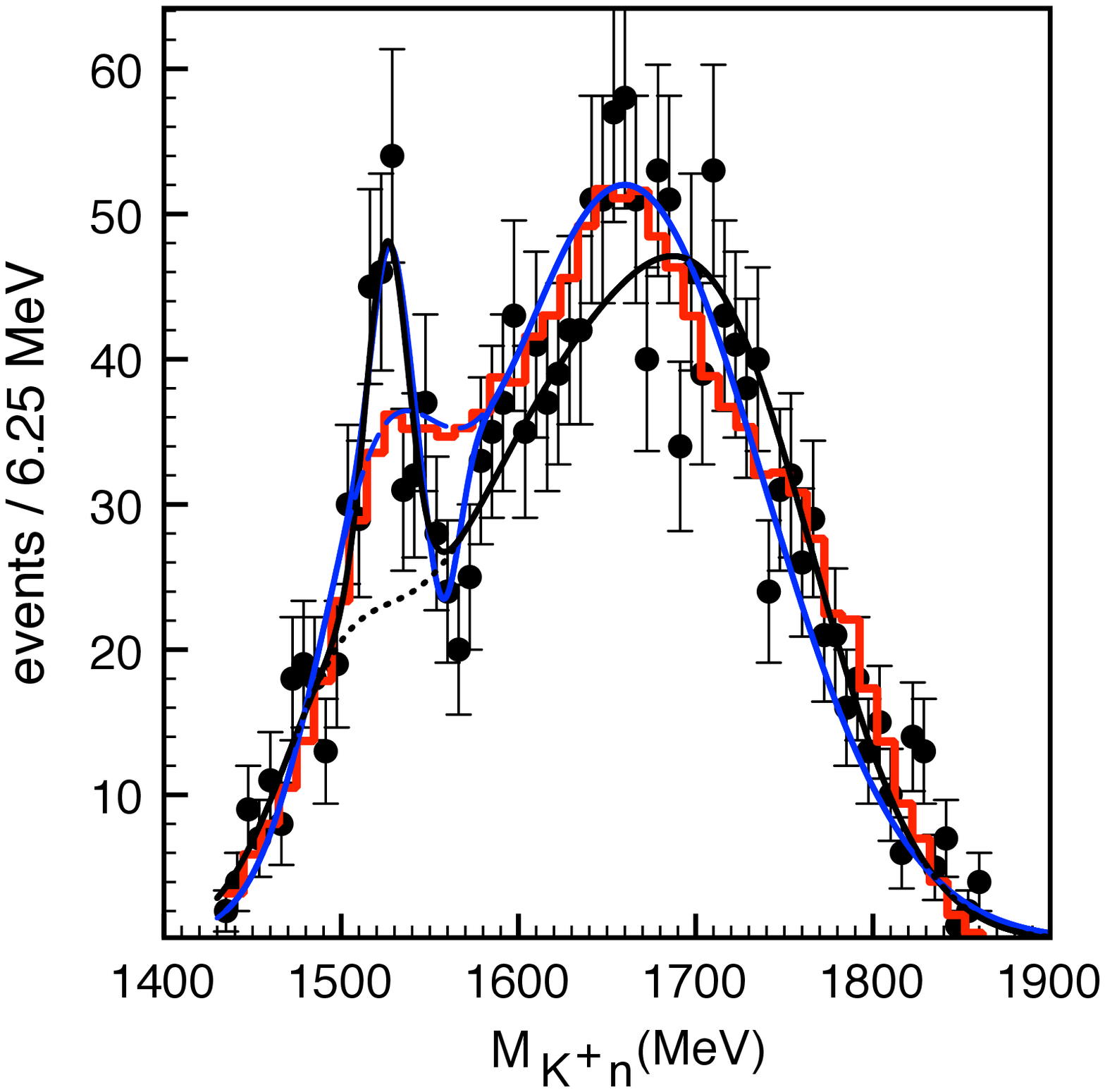}
\caption{(Left) $M_{K^{+}n}$ invariant mass distribution obtained with the MMSA prescription and same cuts than LEPS normalized to the data of \cite{penta2} (shown as dots), together with two gaussians functions: one peaking at 1520 MeV with a width of 80 MeV and another one peaking at 1660 MeV with a width of 185 MeV. (Rigth) The thick black solid line represents the fit of LEPS. The grey solid line (blue online) is the fit with the fluctuation. The dotted line represents  the background of the LEPS fit, while the dashed line (following the histogram) corresponds to  the background of the fit with the fluctuation.}\label{Kmpnorm1}
\end{figure}

The association of the $K^+ N$ spectrum to $K^+ n$ at LEPS has a repercussion in the assumed experimental $K^+ n$ distribution, as shown in  Fig. \ref{Minvnaka_real} (Right) which corresponds to the real distribution obtained within our model. As we can see, there is no peak in that distribution around the ``$\Theta^+$'' peak.  However, in Fig. \ref{Kmpnorm1} (Left) we show the spectrum obtained using the LEPS cuts and the MMSA prescription, normalized to the experimental data. As we can see, the combination of the LEPS cuts and the MMSA prescription has produced an accumulation of strength below  the  ``$\Theta^+$'' region. The shape of the distribution obtained in  Fig. \ref{Kmpnorm1} (Left) can be represented in terms of two gaussians (as shown in the figure), one of them peaking around the ``$\Theta^{+}$" peak. Therefore even in a large statistics experiment one would see this clear broad peak, which could be interpreted as a sign of a resonance. Yet, there is no resonance in that region in the model used.  
 
Here we would like to make a  quantitative study of the statistical significance of the ``$\Theta^+$'' peak. In \cite{penta2} a best fit to the data was done assuming a background and a Gaussian peak in the  ``$\Theta^+$'' region. The best fit with these assumptions provided a background of about 22 events per bin below the ``$\Theta^+$'' peak, as can be seen in Fig. 12 a) of \cite{penta2}. With respect to this background, the ``$\Theta^+$'' peak has a strength of about $5 \sigma$. According to this, the statistical significance of the peak would rule out the possibility of it being a statistical fluctuation. 
 
Conversely, after the theoretical evaluation of the background, our argumentation goes as follows: The actual background below the ``$\Theta^+$'' peak is bigger than the one provided by the LEPS best fit, around 36 events per bin instead of the 22 assumed in \cite{penta2}. This makes the strength of the peak with respect to the background much smaller than in the LEPS best fit. It also makes $\sigma$ larger and the statistical significance is now of about $2 \sigma$, something acceptable as a fluctuation, as in the case of the experimental $K^- p$ spectrum in \cite{penta2}, where a peak is observed around 1650 MeV and associated to a fluctuation of significance $2\sigma$ . This argumentation about the significance of the peaks is corroborated by further calculations which we have carried out: we perform two best fits to the data. One with a background obtained with three broad gaussians and a narrow gaussian below the ``$\Theta^+"$ peak, which gives us a fit practically identical to the one of LEPS. The other one assumes a background and a fluctuation, with a peak and a subsequent dip parametrized as $A[e^{-(x-x_0)^2/\sigma^2}-e^{-(x-x_0-2.3\sigma)^2/\sigma^2}]$. In addition we have two gaussians to account for the background, as in Fig. \ref{Kmpnorm1} (Left). In Fig. \ref{Kmpnorm1} (Right) we show the fit obtained with the two options together with the resulting backgrounds. The $\chi^2$ per degree of freedom  for both cases is 1.2. We observe that in the case of the statistical fluctuation the resulting background is practically identical to the ``exact'' background, while in the case of the LEPS fit the resulting background grossly underestimates the ``exact'' one below the ``$\Theta^+$'' peak.

 In summary, our study has shown that the background in the $\gamma ~d \to ~K^+ K^- ~n ~p $ reaction is fairly larger than the one obtained in the best fit to the data of LEPS assuming a background and a Gaussian peak in the region of the  ``$\Theta^+$''. We also mentioned that the fit of LEPS is not unique and other fits to the data, assuming a background and a fluctuation, are possible, producing the same reduced $\chi^2$ and returning a background nearly identical to the calculated one. Based on the calculated background and the errors obtained from different Monte Carlo runs, we evaluated the statistical significance of the ``$\Theta^+$'' peak and found it to be of about $2 \sigma$ with respect to the background, compatible with a fluctuation. The larger statistical significance claimed in \cite{penta2} was tied to the assumption of a significantly smaller background, which we have found is not justified. \\

\begin{acknowledgments}
The work of A. M. T. is supported by the Grant-in-Aid for the Global COE Program ``The Next Generation of Physics, Spun from Universality and Emergence" from the Ministry of Education, Culture, Sports, Science and Technology (MEXT) of Japan. This work is partly supported by the DGICYT contract FIS2006-03438, the Generalitat Valenciana in the program Prometeo and the EU Integrated Infrastructure Initiative Hadron Physics Project under Grant Agreement n.227431.
  \end{acknowledgments}




\begin{thebibliography}{9}

\bibitem{diakonov}
  D.~Diakonov, V.~Petrov and M.~V.~Polyakov,
  \emph{Z.\ Phys.\  A} {\bf 359}, 305 (1997).

\bibitem{penta1}
 T.~Nakano {\it et al.},
 \emph{Phys.\ Rev.\ Lett.}  {\bf 91}, 012002 (2003).

\bibitem{pdg}
  C.~Amsler {\it et al.},
  \emph{Phys.\ Lett.\  B} {\bf 667}, 1 (2008).
  
\bibitem{penta2}
 T.~Nakano {\it et al.},
 \emph{Phys.\ Rev.\  C} {\bf 79}, 025210 (2009).
 \bibitem{clas}
 B. McKinnon \it{et} al., \emph{Phys. Rev. Lett.} {\bf 96}, 212001 (2006).
\bibitem{MO1} 
A. Mart\'inez Torres and E. Oset, \emph{Phys. Rev. C} {\bf 81}, 055202 (2010);  \emph{Phys.\ Rev.\ Lett.}\  {\bf 105}, 092001 (2010).
\end{thebibliography}
\end{document}